\documentclass[letterpaper]{article}
\usepackage{aaai2026}          
\usepackage{times}
\usepackage{helvet}
\usepackage{courier}
\usepackage[hyphens]{url}
\usepackage{graphicx}
\usepackage{natbib}
\usepackage{caption}
\usepackage{amsmath}
\usepackage{booktabs}
\usepackage{algorithm}
\usepackage{algorithmic}
\usepackage{newfloat}
\usepackage{listings}

\DeclareCaptionStyle{ruled}{labelfont=normalfont,labelsep=colon,strut=off}
\floatstyle{ruled}
\newfloat{listing}{tb}{lst}{}
\floatname{listing}{Listing}

\title{AquaSentinel: Next-Generation AI System Integrating Sensor Networks for Urban Underground Water Pipeline Anomaly Detection via Collaborative MoE-LLM Agent Architecture}

\author{
    Qiming Guo\textsuperscript{1,2}, Bishal Khatri\textsuperscript{1}, Wenbo Sun\textsuperscript{3}, Jinwen Tang\textsuperscript{4}, Hua Zhang\textsuperscript{1}, Wenlu Wang\textsuperscript{1,2}
}
\affiliations{
    \textsuperscript{1}Texas A\&M University - Corpus Christi, USA,\\
    \textsuperscript{2}The AIII Lab, USA,\\
    \textsuperscript{3}Delft University of Technology, Delft, Netherlands,\\
    \textsuperscript{4}University of Missouri, Columbia, MO, USA,\\
    qguo2@islander.tamucc.edu, bkhatri1@islander.tamucc.edu,\\
    w.sun-2@tudelft.nl, jt4cc@umsystem.edu,\\
    hua.zhang@tamucc.edu, wenlu.wang@tamucc.edu
}

\begin{document}

\maketitle

\begin{abstract}
Underground pipeline leaks and infiltrations pose significant threats to water security and environmental safety. Traditional manual inspection methods provide limited coverage and delayed response, often missing critical anomalies. This paper proposes AquaSentinel, a novel physics-informed AI system for real-time anomaly detection in urban underground water pipeline networks. We introduce four key innovations: (1) strategic sparse sensor deployment at high-centrality nodes combined with physics-based state augmentation to achieve network-wide observability from minimal infrastructure; (2) the RTCA (Real-Time Cumulative Anomaly) detection algorithm, which employs dual-threshold monitoring with adaptive statistics to distinguish transient fluctuations from genuine anomalies; (3) a Mixture of Experts (MoE) ensemble of spatiotemporal graph neural networks that provides robust predictions by dynamically weighting model contributions; (4) causal flow-based leak localization that traces anomalies upstream to identify source nodes and affected pipe segments. Our system strategically deploys sensors at critical network junctions and leverages physics-based modeling to propagate measurements to unmonitored nodes, creating virtual sensors that enhance data availability across the entire network. Experimental evaluation using 110 leak scenarios demonstrates that AquaSentinel achieves 100\% detection accuracy. This work advances pipeline monitoring by demonstrating that physics-informed sparse sensing can match the performance of dense deployments at a fraction of the cost, providing a practical solution for aging urban infrastructure.
\end{abstract}

\begin{links}
    \link{Dataset}{https://github.com/VV123/STEPS}
\end{links}

\section{Introduction}

Underground water pipelines are the hidden lifelines of modern cities, ensuring continuous delivery of clean water and supporting urban resilience. Yet leaks and infiltrations remain widespread and costly: in the United States alone, over 2.1 trillion gallons of treated water are lost each year to undetected anomalies~\cite{EPA2020}. Beyond the economic waste, such failures trigger cascading risks to public health, environmental safety, and infrastructure stability, underscoring the urgent need for monitoring systems that can detect anomalies in real time, before minor leaks escalate into catastrophic failures.

\textbf{Motivating Scenario}. Consider a dense metropolitan water network with thousands of underground junctions. Subtle pressure drops in one buried segment may signal the onset of corrosion-induced leakage. Without timely detection, background leaks silently undermine soil integrity and groundwater quality, ultimately manifesting as visible ruptures that flood streets and disrupt critical services. Because urban underground pipelines are deeply buried, conducting frequent manual inspections across the entire network is practically infeasible. Neither the required manpower nor the financial resources are proportionate to the benefits, limiting manual inspections to only a few critical junctions. Likewise, dense sensor networks—while theoretically ideal—remain economically prohibitive for legacy infrastructure, constraining comprehensive monitoring to newly built or small-scale areas.

Recent major water infrastructure failures in the United States underscore the urgent need for proactive monitoring. In Houston (2024), a 96-inch water main rupture triggered a city-wide boil-water advisory and flooded major freeways~\cite{case15}; in Detroit (2023), a large break affected nearly 400 homes, forcing evacuations~\cite{case22}; and in Carnelian Bay, California (2021), a sewer spill of tens of thousands of gallons closed beaches and resulted in an \$850,000 settlement~\cite{case21}. 

\begin{figure*}[h!]
    \centering
    \includegraphics[width=0.96\textwidth]{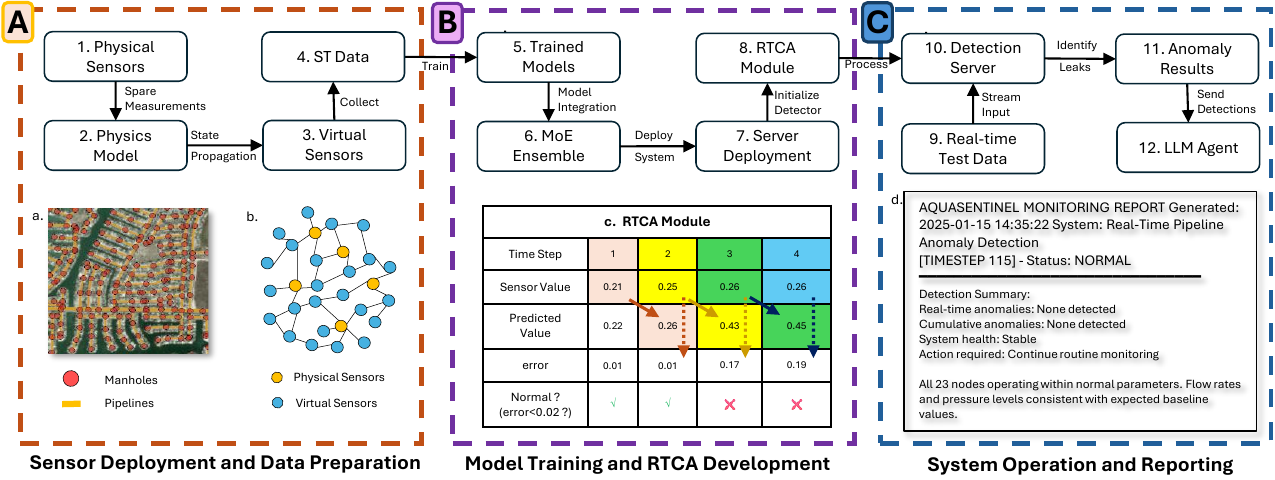}
    \caption{AquaSentinel end-to-end architecture overview. \textbf{(A)} Sparse sensor deployment and physics-based virtual sensor creation; \textbf{(B)} MoE ensemble of spatiotemporal GNNs with RTCA dual-threshold anomaly detection; \textbf{(C)} Causal localization and LLM-generated actionable reports for maintenance dispatch.}
    \label{fig:system_architecture}
\end{figure*}

These incidents represent two typical failure modes: (1) \textbf{background leakage}—small, persistent leaks at joints or due to corrosion that remain hidden underground for months or years, silently wasting water, contaminating soil and groundwater, and weakening structural integrity; and (2) \textbf{visible (catastrophic) leakage}—sudden, surface-manifested ruptures that disrupt services and cause immediate damage. Critically, many visible failures originate from undetected background leaks that progressively worsen over time. Early detection of background leakage is therefore essential to prevent escalation into costly disasters, yet current methods remain either too expensive for widespread deployment or insufficiently sensitive to subtle anomalies.

\textbf{Research Gap}. Existing approaches face a fundamental trade-off: dense sensor deployments provide comprehensive coverage but incur prohibitive costs, while sparse deployments reduce costs but sacrifice detection capability. Traditional methods assume that effective monitoring requires sensors at most or all network nodes—an assumption that renders retrofitting of legacy infrastructure economically infeasible due to linearly scaling installation costs. Moreover, purely data-driven methods struggle with limited observations, physics-only models fail to adapt to changing network conditions, and no prior work has yet achieved network-wide anomaly detection from sparse sensors through tight integration of physical laws and learning-based prediction.

Recent advances in LLM-based agents have dramatically enhanced robustness, automation, and cross-domain reasoning in fields ranging from civil engineering~\cite{liang2025integrating,goldshtein2025large} and hydrological time-series forecasting~\cite{qin2025llm} to education~\cite{tang2025tigergpt} and psychology~\cite{guo2024soullmate}.AquaSentinel draws inspiration from these trends by combining sparse sensor data with physics-informed predictions, MoE ensembles for robustness, and an LLM agent for interpretable reporting—realizing an end-to-end system for underground infrastructure that was previously infeasible.

In this study, we propose \textbf{AquaSentinel}, a physics-informed AI framework that fundamentally challenges the assumption that comprehensive monitoring requires dense sensor coverage. Our key insight is that hydraulic networks exhibit strong spatial correlations governed by physical laws—anomalies at unmonitored locations propagate predictable effects to sensor locations. By combining strategic sensor placement, physics-based state augmentation, and spatiotemporal learning, AquaSentinel achieves network-wide anomaly detection using only 20-30\% sensor coverage. The system comprises: (i) strategic deployment at high-centrality nodes to maximize information flow, (ii) physics-based augmentation using conservation laws to infer states at unmonitored nodes, (iii) the Real-Time Cumulative Anomaly (RTCA) algorithm with dual-threshold detection, and (iv) causal flow analysis for precise leak localization.

\textbf{Challenge 1: Achieving comprehensive coverage with sparse sensors.}  
How can utilities monitor entire networks when sensors cover only a fraction of nodes?

\textbf{Solution:} AquaSentinel leverages physics-based state propagation to create "virtual sensors" at unmonitored nodes, using conservation laws to infer network-wide states from sparse measurements.

\medskip
\textbf{Challenge 2: Robust detection despite incomplete observations.}  
How can systems distinguish genuine anomalies from noise when observing limited nodes?

\textbf{Solution:} AquaSentinel's RTCA algorithm employs dual-threshold monitoring with adaptive statistics, requiring both instantaneous and cumulative deviations to persist across multiple timesteps before confirming anomalies.

\medskip
\textbf{Challenge 3: Precise localization from indirect observations.}  
How can the exact leak location be identified when the anomaly occurs at an unmonitored node?

\textbf{Solution:} AquaSentinel exploits flow causality, tracing detected anomalies upstream to identify source nodes, with leak segments localized between sources and their nearest normal upstream neighbors.

\textbf{Contributions.} First, we demonstrate that \textbf{sparse sensor deployment with physics augmentation} can achieve comprehensive network monitoring, challenging the traditional dense-coverage paradigm. Second, we propose the \textbf{RTCA detection framework} that maintains high sensitivity while suppressing false positives through dual-threshold adaptive monitoring. Third, we introduce \textbf{causal flow-based localization} that precisely identifies leak sources by exploiting hydraulic network topology. Validated on PCSWMM-simulated datasets calibrated with real sensor data, AquaSentinel achieves \textbf{100\% detection accuracy across 110 leakage cases, with 90.91\% of anomalies detected within 10 timesteps}. This work establishes a new paradigm for infrastructure monitoring: comprehensive awareness through intelligent sparse sensing.

\section{AquaSentinel Principles}

We propose AquaSentinel, a physics-informed AI framework achieving comprehensive pipeline anomaly detection through sparse sensor deployment. The system orchestrates strategic sensor placement, physics-based state augmentation, spatiotemporal prediction, and our novel RTCA (Real-Time Cumulative Anomaly) detection algorithm. Unlike traditional approaches requiring dense coverage, AquaSentinel achieves network-wide monitoring using only 20-30\% sensor coverage by leveraging hydraulic constraints and spatiotemporal learning.

\subsection{Problem Formulation}

Consider a pipeline network $G = (V, E)$ with $|V| = n$ nodes and directed edges representing flow. Each node $v \in V$ has time-varying states $\mathbf{x}_v(t) = [q_v(t), h_v(t), p_v(t)]^T$ denoting flow rate, depth, and pressure. Given partial observations $\mathcal{S} \subset V$ where $|\mathcal{S}| \ll |V|$, we aim to detect and localize anomalies across the entire network:

\begin{equation}
\text{Detect: } \mathcal{A}(t) = \{v \in V : \|\mathbf{x}_v(t) - \hat{\mathbf{x}}_v(t)\| > \tau(v, t)\}
\end{equation}

\begin{equation}
\text{Localize: } v^* = \arg\min_{v \in \mathcal{A}(t)} |\text{Upstream}(v) \cap \mathcal{A}(t)|
\end{equation}

where $\hat{\mathbf{x}}_v(t)$ is the predicted normal state and $\tau(v, t)$ are adaptive thresholds. The challenge is achieving $P(\text{detect}|\text{anomaly}) > 0.95$ with $P(\text{false alarm}) < 0.01$ from sparse observations, while precisely identifying leak locations for maintenance dispatch.

\begin{figure*}[t!]
    \centering
    \includegraphics[width=0.96\textwidth]{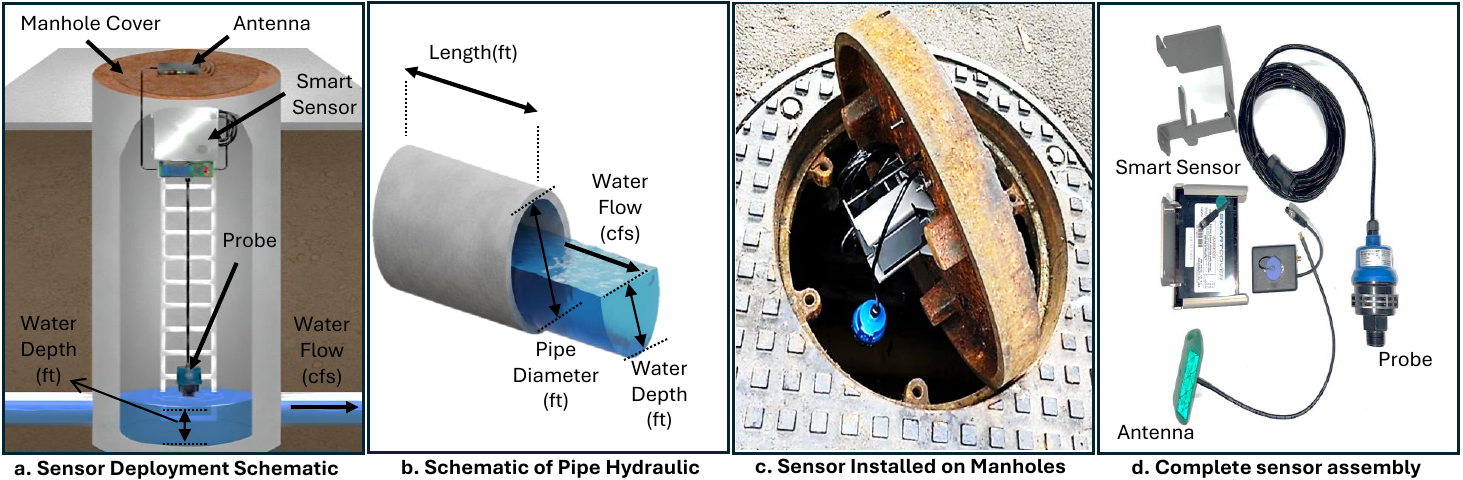}
    \caption{AquaSentinel sparse sensor deployment and hardware implementation.}
    \label{fig:hardware}
\end{figure*}

\subsection{Strategic Sensor Deployment}

Optimal sensor placement maximizes observability while minimizing deployment cost. Rather than uniform coverage, we strategically position sensors at high-impact nodes. We compute node importance by combining topological centrality, hydraulic significance, and risk factors:

\begin{equation}
\text{Score}(v) = \alpha \cdot C_B(v) + \beta \cdot \text{Hydraulic}(v) + \gamma \cdot \text{Risk}(v)
\end{equation}

where $C_B(v)$ is betweenness centrality identifying critical flow paths, 
$\text{Hydraulic}(v) = \bar{Q}_v \cdot \Delta P_v$ captures flow volume and pressure dynamics, 
and $\text{Risk}(v)$ incorporates historical failure rates and infrastructure age.

The final sensor selection problem is then formulated as:

\begin{equation}
\begin{aligned}
\mathcal{S}^* = \arg\max_{\mathcal{S} \subset V,\, |\mathcal{S}| \leq k} 
    & \sum_{v \in \mathcal{S}} \text{Score}(v) \\
\text{subject to} \quad & d(u,v) \geq d_{\min}, \quad \forall u,v \in \mathcal{S}
\end{aligned}
\end{equation}

ensuring both high information gain and uniform spatial distribution to avoid blind spots.

\subsection{Physics-Based State Augmentation}

Physical conservation laws enable state estimation at unmonitored nodes, effectively creating "virtual sensors" throughout the network. Given sensor readings $\mathbf{X}_{\mathcal{S}}(t)$, we estimate states at remaining nodes through physics-constrained optimization:

\begin{equation}
\mathbf{X}_{V \setminus \mathcal{S}}^* = \arg\min_{\mathbf{X}} \|\mathcal{F}(\mathbf{X})\|^2 + \lambda\|\nabla\mathbf{X}\|^2
\end{equation}

The physics constraints $\mathcal{F}$ enforce mass conservation at junctions:
\begin{equation}
\sum_{i \in \text{In}(v)} Q_i = \sum_{j \in \text{Out}(v)} Q_j + D_v, \quad \forall v \in V
\end{equation}

and energy conservation along pipes via Hazen-Williams:
\begin{equation}
h_f = 10.67L \left(\frac{Q}{CD^{2.63}}\right)^{1.852}
\end{equation}

where $h_f$ is head loss, $L$ is length, $C$ is roughness coefficient, and $D$ is diameter. This augmentation propagates sensor observations throughout the network, amplifying anomaly signals beyond direct measurement points.









\subsection{Spatiotemporal Prediction via MoE}
Accurate anomaly detection relies on robust forecasting of normal hydraulic states from sparse observations. We employ a Mixture-of-Experts (MoE) ensemble with six recent spatiotemporal models as experts: CaST~\cite{xia2024deciphering}, GMAN~\cite{zheng2020gman}, ST-SSL~\cite{ji2023spatio}, STG-MAMBA~\cite{li2024stg}, STGCN~\cite{yu2018spatio}, and HydroNet~\cite{guo2025ai}. As shown in Table~\ref{tab:rww_results}, these models exhibit complementary strengths on the real-world campus dataset, with STGCN and HydroNet consistently ranking highest.

The MoE gating network dynamically weights the experts using exponentially smoothed recent loss:
\begin{equation}
\begin{aligned}
\hat{\mathbf{y}}_t &= \sum_{m=1}^{6} w_m(t) \cdot f_m(\mathbf{X}_{t-T:t}, \mathbf{A}), \\[4pt]
w_m(t) &= \frac{\exp(-\lambda \mathcal{L}_m(t))}{\sum_{j=1}^{6} \exp(-\lambda \mathcal{L}_j(t))}.
\end{aligned}
\end{equation}
This adaptive mechanism ensures stable, high-quality predictions even under sensor noise and varying flow conditions.

\subsection{RTCA: Real-Time Cumulative Anomaly Detection}

RTCA resolves the sensitivity-specificity trade-off through dual-threshold monitoring. For each node, we track instantaneous and cumulative deviations.

Real-time error captures immediate anomalies:
\begin{equation}
e_t^{\text{RT}}(v) = \frac{|y_t(v) - \hat{y}_t(v)|}{\hat{y}_t(v) + \epsilon}
\end{equation}

Cumulative error aggregates recent history over window $W$:
\begin{equation}
e_t^{\text{C}}(v) = \frac{1}{W}\sum_{i=t-W+1}^{t} e_i^{\text{RT}}(v)
\end{equation}

Adaptive thresholds evolve via exponential moving average:
\begin{align}
\mu_t(v) &= (1-\alpha)\mu_{t-1}(v) + \alpha e_t^{\text{RT}}(v) \\
\sigma_t^2(v) &= (1-\alpha)\sigma_{t-1}^2(v) + \alpha(e_t^{\text{RT}}(v) - \mu_t(v))^2 \\
\tau^{\text{RT}}_t &= \mu_t + k_1\sigma_t, \quad \tau^{\text{C}}_t = \mu_t + k_2\sigma_t
\end{align}

where $k_1 = 2.5$ and $k_2 = 3.0$ balance sensitivity and specificity.

Anomaly confirmation requires both thresholds exceeded for $T$ consecutive steps:
\begin{equation}
\mathcal{A}(v,t) = 1\left[\bigwedge_{\tau=t-T+1}^{t} (e_\tau^{\text{RT}} > \tau^{\text{RT}}_\tau \land e_\tau^{\text{C}} > \tau^{\text{C}}_\tau)\right]
\end{equation}

Upon detection, we trace upstream to localize the source:
\begin{equation}
v^* = \{v \in \mathcal{A} : \text{Upstream}(v) \cap \mathcal{A} = \emptyset\}
\end{equation}

The leak segment lies between $v^*$ and its nearest normal upstream neighbor.

\subsection{Intelligent Report Generation}

A domain-specific LLM translates algorithmic outputs into actionable maintenance instructions. The system processes detection results through structured prompting:

\begin{equation}
\mathcal{R} = \text{LLM}(\mathcal{D}, \mathcal{C}, \mathcal{H}, \mathcal{T})
\end{equation}

where $\mathcal{D}$ contains detection results, $\mathcal{C}$ provides network context, $\mathcal{H}$ includes historical patterns, and $\mathcal{T}$ specifies report templates.

Severity classification determines response urgency:
\begin{equation}
\text{Severity}(v) = \begin{cases}
\text{Critical} & \text{if } \text{Conf}(v) > 0.9 \land e^{\text{RT}} > 0.3 \\
\text{Major} & \text{if } \text{Conf}(v) > 0.7 \land e^{\text{RT}} > 0.15 \\
\text{Minor} & \text{otherwise}
\end{cases}
\end{equation}

Maintenance prioritization combines multiple factors:
\begin{equation}
\text{Priority}(v) = \text{Conf}(v) \cdot C_B(v) \cdot \text{Impact}(v)
\end{equation}

The generated report includes executive summary, technical details, resource requirements, and safety protocols, bridging the gap between AI precision and field operations.

\textbf{System Integration}. AquaSentinel operates as an integrated pipeline where each component reinforces the others. Strategic sensor deployment provides critical observations at network bottlenecks. Physics-based augmentation extends these sparse measurements across the entire network, creating comprehensive state awareness. The MoE ensemble learns normal patterns from this augmented data, providing robust predictions despite model uncertainties. RTCA's dual-threshold mechanism maintains high sensitivity while filtering transient noise, confirmed through persistence requirements. Finally, causal flow analysis enables precise leak localization, while LLM-generated reports ensure rapid, appropriate field response. This synergistic design achieves what neither pure physics models nor data-driven approaches can accomplish alone: comprehensive, accurate, and actionable anomaly detection from minimal sensor infrastructure, fundamentally transforming how urban water networks are monitored and maintained.

\section{AquaSentinel Deployment and Evaluation}

To evaluate whether \textbf{AquaSentinel} is a reliable approach, we address five key questions:  
1. Can AquaSentinel’s concept of deploying only a subset of sensors be effective?  
2. Are graph-structured spatiotemporal data suitable sources for supporting anomaly detection?  
3. Is relying solely on manhole-installed sensors sufficient for accurate anomaly localization?  
4. Can AquaSentinel detect both short-term anomalies and long-term cumulative anomalies?  
5. Can the LLM-based reporting system correctly interpret monitoring data and generate actionable reports?  

To answer these questions, we deployed AquaSentinel in a real-world environment. The details of the deployment are described below.

\textbf{Test Site.} Sensors were installed on manholes in an underground sewer pipeline network located beneath the Texas A\&M University--Corpus Christi campus. The sewer system was under normal operation and maintenance during deployment, and no leaks were present, ensuring that experimental anomalies could be clearly distinguished from natural conditions.  

\textbf{Sensor Deployment.} Given AquaSentinel’s goal of reducing costs and avoiding invasive retrofitting of existing pipelines, we adopted a minimally intrusive deployment strategy. Sensors were mounted on the back side of manholes, which are naturally present in sewer networks. This placement does not interfere with manhole usage and avoids requiring human entry into pipelines for installation.  

To minimize system cost, each sensor collected only two features: flow velocity and water depth. To reduce labor costs and eliminate the need for manual data retrieval, the sensors were designed for real-time network transmission of collected data. Each device was battery-powered with an expected lifetime of \textbf{two years}, enabling long-term continuous operation.

\textbf{Data Collection and Processing.}
Monitoring data were transmitted to a Google Cloud Platform (GCP) server, with local storage available at the sensor nodes for redundancy. Data collection began on October 1, 2023, with a temporal resolution of 10 minutes per timestep (six samples per hour). The recorded features included flow velocity and water depth at each instrumented manhole.  

To achieve network-wide coverage, the observed data were integrated into a \textbf{PCSWMM} hydraulic model of the sewer network, which simulated flow velocity and water depth for all 23 nodes at the same temporal resolution. These augmented datasets were used to train spatiotemporal prediction models such as STGCN and Graph WaveNet, which were then combined in a Mixture of Experts (MoE) framework. For any 12-step sequence of observations across the 23 nodes, the models were trained to forecast either the next single timestep or a sequence of 12 future timesteps. The real-world water dataset is publicly available~\url{https://github.com/VV123/STEPS} and has supported benchmarking in prior work~\cite{guo2025efficient}~\cite{guo2025ai}~\cite{guo2024hydronet}~\cite{guo2025spatio}.

\textbf{Data Quality and Model Accuracy.} To validate the quality of real-world sensor data and select strong MoE experts, we evaluated six recent spatiotemporal forecasting models on water depth and flow rate prediction using MAE and RMSE metrics (Table~\ref{tab:rww_results}). The consistently low errors across models confirm excellent data quality, while STGCN and HydroNet emerge as top performers, justifying their dominant weighting in the MoE ensemble.

\textbf{Anomaly Detection and Localization Principle.}  
As described in Section \textit{RTCA: Real-Time Cumulative Anomaly Detection}, AquaSentinel monitors both short-term anomalies and cumulative anomalies. The process operates as follows:  

1. Models are trained on long-term normal operation data from the deployed sensors to construct the MoE system.  

2. At runtime, for each new timestep, the system uses the previous 11 timesteps plus the current sensor reading to form a 12-step input window, and predicts the 13th timestep.  

3. When the actual sensor reading for the 13th timestep becomes available (10 minutes later), it is compared against the predicted value. If the absolute error is below the predefined threshold, the system continues monitoring and marks the timestep as normal. Each absolute error is also accumulated into a cumulative error score.  
   
4. As long as both instantaneous and cumulative errors remain below their respective thresholds, the system does not raise an alert.  

5. If thresholds are exceeded persistently, the system confirms the anomaly and activates causal flow-based localization to identify the source.  

This mechanism allows AquaSentinel to simultaneously detect transient anomalies and long-term degradations while suppressing false positives.

\begin{table}[t!]
\caption{Prediction performance of MoE expert submodels on the real-world campus sewer dataset (lower is better; best per column in bold).}
\label{tab:rww_results}
\centering
\footnotesize
\setlength{\tabcolsep}{8pt}
\renewcommand{\arraystretch}{1.08}
\begin{tabular}{l cc cc}
\toprule
\textbf{Method} & \multicolumn{2}{c}{\textbf{Depth (ft)}} & \multicolumn{2}{c}{\textbf{Flow (cfs)}} \\
\cmidrule(lr){2-3} \cmidrule(lr){4-5}
 & MAE & RMSE & MAE & RMSE \\
\midrule
CaST       & 0.0186 & 0.0298 & 0.0077 & 0.0138 \\
GMAN       & 0.0186 & 0.0140 & 0.0168 & 0.0181 \\
ST-SSL     & 0.0196 & 0.0230 & 0.0150 & 0.0322 \\
STG-MAMBA  & 0.0176 & 0.0296 & 0.0098 & 0.0166 \\
STGCN      & \textbf{0.0123} & 0.0324 & \textbf{0.0066} & 0.0158 \\
HydroNet   & 0.0085 & \textbf{0.0178} & 0.0038 & \textbf{0.0094} \\
\bottomrule
\end{tabular}
\end{table}

\begin{table*}[t]
\centering
\caption{Leak detection results across different scenarios (22 conduits $\times$ 5 leakage types = 110 cases). Detection delay is measured in timesteps (10 minutes per step). All 110 leaks were eventually detected.}
\label{tab:leakage_detection}
\begin{tabular}{l c c c c}
\toprule
\textbf{Leakage Scenario} &
\textbf{Detected} &
\textbf{Avg. Detection} &
\textbf{Proportion Detected} &
\textbf{Overall} \\
&
&
\textbf{Delay (timesteps)} &
\textbf{within 10 timesteps} &
\textbf{Detection Rate} \\
\midrule
Constant $<$5\%                     & Yes & 12 & 81.8\%  & 100\% \\
Constant 5--15\%                    & Yes &  8 & 100\%   & 100\% \\
Constant 15--25\%                   & Yes &  5 & 100\%   & 100\% \\
Constant $>$25\%                    & Yes &  3 & 100\%   & 100\% \\
Dynamic Leakage                     & Yes & 15 & 73.6\%  & 100\% \\
(increasing from 0\% to 35\%)       &     &    &         &       \\
\midrule
\textbf{Overall (110 cases)}        & Yes & -- & 90.91\% & 100\% \\
\bottomrule
\end{tabular}
\end{table*}

\textbf{Evaluation Data Preparation.} A robust dataset representing both normal and anomalous network states was synthetically generated to facilitate the development and evaluation of the proposed AquaSentinel. The foundation of this process was a high-fidelity hydraulic model of the Texas A\&M University-Corpus Christi (TAMUCC) campus sewer network, developed within the Personal Computer Stromwater Management Model (PCSWMM) environment. To ensure the model accurately replicated real-world system dynamics, it was calibrated using flow data acquired from a network of Smartcover  sensors deployed on campus.

First, a baseline simulation was executed for a duration of one week with a temporal resolution of 10 minutes. This simulation, which was devoid of any induced faults, produced a reference dataset representing the normal flow characteristic of sewer network.

Subsequently, leakage events were systematically introduced into the model's 22 conduits. The simulation of these leaks was accomplished by leveraging specific hydraulic modeling elements within PCSWMM, using two distinct methods: 

\begin{enumerate}
    \item \textbf{Constant Leakage:} To model a constant leak, an Outlet link element was added to the target conduit. This outlet was configured to discharge to a free outfall node, simulating flow leaving the system. The discharge coefficient of the outlet was then precisely calibrated to produce a steady leakage rate. Four scenarios were simulated a constant leakage rate. The magnitude of these leaks was defined as a percentage of the flow at the conduit's upstream node, categorized as: $<$5\%, 5\%--15\%, 15\%--25\%, and $>$25\%.

    \item \textbf{Dynamic Leakage:} A fifth scenario simulated a dynamic leakage event. This was characterized by a rate that incrementally increased from a low initial value (approximately 0.5\%--1\%) to a predefined maximum before stabilizing for the remainder of the duration. To model a leak with a variable rate, a time series file containing the desired outflow pattern, starting small and increasing to a maximum, was created. In the model, the target conduit was split to introduce a new junction node at the point of the leak. The time series was then assigned to this new node as a direct inflow, with negative values representing water leaving the system. 

\end{enumerate}

Each of the 110 leakage simulations (22 conduits $\times$ 5 scenarios) maintained the same one-week duration and 10-minute time step as the baseline. The result of this process is a synthetic, high-resolution dataset that encompasses a wide spectrum of leakage magnitudes and temporal dynamics across all network segments. This dataset provides the necessary foundation for the performance assessment of the proposed leakage detection.

\textbf{The evaluation results} see Table~\ref{tab:leakage_detection}) demonstrate that AquaSentinel effectively detected anomalies across all 110 simulated leakage cases, achieving \textbf{100\% detection accuracy}. Clearly, constant large-value leakages are easier to detect within only 3–5 timesteps, whereas constant small-flow leakages require more timesteps (up to 12) to be identified due to the cumulative error effect. For dynamic leakage, because the outflow gradually increases from 0\% to 35\%, it is detected after approximately 15 timesteps, showing a detection delay comparable to that of the Constant < 5\% leakage scenario. Overall, \textbf{90.91\% of all leakages were successfully detected within 10 timesteps}.

Also, from the evaluation results in Table~\ref{tab:rww_results} and Table~\ref{tab:leakage_detection}, the following key questions can be answered. 

First, AquaSentinel’s concept of partial sensor deployment is effective: even with sparse placement, physics-informed propagation yields data of comparable utility to real sensors.  

Second, graph-structured spatiotemporal data are indeed suitable for anomaly detection, as confirmed by the high accuracy in Table~\ref{tab:rww_results}.  

Third, sensors mounted on manholes are easy to install and reliably collect data without invasive pipeline entry.  

Fourth, AquaSentinel successfully detects both transient and cumulative anomalies—RTCA captures small leaks through accumulated error and large leaks within only a few timesteps.  

Finally, the LLM-based reporting system can transform time-series data and anomaly flags into actionable reports, supporting field decision-making.  

\section{Limitations}
Validation relies primarily on high-fidelity simulations, as no confirmed real-world leaks occurred during the study period. Full-scale municipal deployments are planned to verify performance under real conditions.

\section{Conclusion}
This paper presented \textbf{AquaSentinel}, a physics-informed AI framework for real-time anomaly detection in urban underground water pipelines. By integrating sparse physics-augmented sensing, a Mixture-of-Experts ensemble of spatiotemporal GNNs, the RTCA dual-threshold detector, and a domain-specific LLM agent for interpretable reporting, AquaSentinel delivers an end-to-end solution that surpasses prior manual, sensor-based, and AI-only approaches.

Evaluation on 110 PCSWMM-simulated leakage scenarios shows \textbf{100\% detection accuracy}, with \textbf{90.91\%} of anomalies identified within 10 timesteps (100 minutes). These results resolve the long-standing trilemma of cost, robustness, and interpretability.

Future work includes city-scale municipal deployments, controlled real-leak field trials for threshold refinement, and multi-modal sensor fusion with advanced architectures to enable even earlier detection.

\section{Acknowledgments}
This work is supported by NSF awards No. 2318641.

\clearpage
\section{Ethics Statement}

In this research, we developed the AquaSentinel framework for anomaly detection in urban water pipeline networks. Our experiments utilized two datasets: (i) a synthetic dataset generated using the PCSWMM hydraulic modeling environment, and (ii) a real-world sewer monitoring dataset collected on the Texas A\&M University--Corpus Christi campus with proper authorization. The collected data include only flow velocity and water depth measurements and contain no personally identifiable information. All privacy and safety considerations were duly addressed during data collection and analysis.

\bibliography{aaai2026}

\end{document}